\documentclass[aps,prx,twocolumn,showpacs,superscriptaddress,longbibliography]{revtex4-2}
\usepackage{amsmath,amssymb,graphicx}
\usepackage{graphicx,epstopdf}
\usepackage{gensymb}
\usepackage[dvipsnames]{xcolor}
\usepackage[T1]{fontenc}
\usepackage[utf8]{inputenc}
\newcommand{\be}{\begin{equation}}
\newcommand{\ee}{\end{equation}}
\newcommand{\bea}{\begin{eqnarray}}
\newcommand{\eea}{\end{eqnarray}}
\newcommand{\bse}{\begin{subequations}}
\newcommand{\ese}{\end{subequations}}

\usepackage{multibib}
\usepackage{color}
\usepackage[colorlinks,bookmarks=false,citecolor=darkblue,linkcolor=red,urlcolor=blue]{hyperref}

\definecolor{darkred}{rgb}{0.7,0.0,0.0}

\definecolor{darkblue}{rgb}{0,0.02,0.45}

\definecolor{darkgreen}{rgb}{0.02,0.45,0.0}

\definecolor{violet}{rgb}{0.8,0.2,0.6}

\begin{document}

\title{Ground state properties of a spin-$\frac{5}{2}$ frustrated triangular lattice antiferromagnet NH$_{4}$Fe(PO$_{3}$F)$_2$}

\author{S. Mohanty}
\affiliation{School of Physics, Indian Institute of Science Education and Research Thiruvananthapuram 695551, India}
\author{K. M. Ranjith}
\affiliation{Leibniz Institute for Solid State and Materials Research Dresden, 01069 Dresden, Germany}
\affiliation{Würzburg-Dresden Cluster of Excellence ct.qmat, Technische Universität Dresden, 01069 Dresden, Germany}
\author{C. S. Saramgi}
\affiliation{Leibniz Institute for Solid State and Materials Research Dresden, 01069 Dresden, Germany}
\author{Y. Skourski}
\affiliation{High Magnetic Field Laboratory (HLD-EMFL), Helmholtz-Zentrum Dresden-Rossendorf, 01328 Dresden, Germany}
\author{B. B\"uchner}
\affiliation{Leibniz Institute for Solid State and Materials Research Dresden, 01069 Dresden, Germany}
\affiliation{Würzburg-Dresden Cluster of Excellence ct.qmat, Technische Universität Dresden, 01069 Dresden, Germany}
\author{H. -J. Grafe}
\affiliation{Leibniz Institute for Solid State and Materials Research Dresden, 01069 Dresden, Germany}
\author{R. Nath}
\email{rnath@iisertvm.ac.in}
\affiliation{School of Physics, Indian Institute of Science Education and Research Thiruvananthapuram 695551, India}

\date{\today}

\begin{abstract}
Structural and magnetic properties of a two-dimensional spin-$\frac{5}{2}$ frustrated triangular lattice antiferromagnet NH$_{4}$Fe(PO$_{3}$F)$_2$ are explored via x-ray diffraction, magnetic susceptibility, high-field magnetization, heat capacity, and $^{31}$P nuclear magnetic resonance experiments on a polycrystalline sample. The compound portrays distorted triangular units of the Fe$^{3+}$ ions with anisotropic bond lengths.
The magnetic susceptibility shows a broad maxima around $T^{\rm{max}}_{\chi}\simeq 12$~K, mimicking the short-range antiferromagnetic order of a low-dimensional spin system. The magnetic susceptibility and NMR shift could be modeled assuming the spin-$5/2$ isotropic triangular lattice model and the average value of the exchange coupling is estimated to be $J/k_{\rm B} \simeq 1.7$~K. This value of the exchange coupling is reproduced well from the saturation field of the pulse field data. It shows the onset of a magnetic ordering at $T_{\rm N} \simeq 5.7$~K, setting the frustration ratio of $f = \frac{|\theta_{\rm CW}|}{T_{\rm N}} \simeq 5.7$. Such a value of $f$ reflects moderate magnetic frustration in the compound. The d$M$/d$H$ vs $H$ plots of the low temperature magnetic isotherms exhibit a sharp peak at $H_{\rm SF} \simeq 1.45$~T, suggesting a field-induced spin-flop transition and magnetic anisotropy. The rectangular shape of the $^{31}$P NMR spectra below $T_{\rm N}$ unfolds that the ordering is commensurate antiferromagnet type. Three distinct phase regimes are clearly discerned in the $H - T$ phase diagram, redolent of a frustrated magnet with in-plane (XY-type) anisotropy.
\end{abstract}

\maketitle

\section{Introduction} 
Frustrated and low-dimensional spin systems have attracted enormous attention in recent days because of their potential to host unusual ground states and low temperature properties~\cite{Diep2013}. Among the broad family of the geometrically frustrated magnets, the two-dimensional (2D) triangular lattice antiferromagnets (TLAFs) show rich physics with varieties of exotic states depending on the temperature and magnetic field~\cite{Collins605,*Starykh052502}. One of the most notable examples is the quantum spin liquid (QSL), a highly entangled and disordered state with no magnetic long-range-order (LRO). For a spin-$1/2$ Heisenberg TLAF, Anderson predicted the resonating-valance-bond state, which is a prototype of QSL state~\cite{Anderson153,*Savary016502}.
Apart from QSL, non-collinear 120$\degree$, collinear antiferromagnetic, and other atypical states are also envisaged theoretically and later perceived experimentally~\cite{Capriotti3899,*Zhu207203,BhattacharyaL060403,Chernyshev144416,Somesh104422,Sebastian104425,Smirnov037202,Ranjith094426}.

In isotropic Heisenberg TLAFs with uniform nearest neighbor (NN) exchange, the ground state is predicted to be a 120$\degree$ non-collinear state for spin, $S > 1/2$~\cite{Jolicoeur2727,Bernu2590,Rawl054412}. Later, using the spin wave theory it was proposed that even for $S=1/2$, the ground state should be non-collinear but with a reduction in the sub-lattice magnetization ~\cite{Jolicoeur2727}.
However, in real materials, the crystal structures are often distorted, leading to non-uniform bond lengths along the sides of the triangle. In TLAFs with isosceles type distortion (with two unequal exchange interactions, $J$ and $J^{\prime}$), the effect of frustration is enhanced and the nature of the ground state is strongly dependent on the relative strength ($J^{\prime}$/$J$) of these interactions~\cite{Yunoki014408,Weng012407}.
Similarly, the ground state also varies drastically when the interactions beyond NN ($J_1$) are taken into consideration. For a significant next nearest neighbor (NNN) interaction ($J_2$), a phase diagram has been reported theoretically depending on the value of $J_2/J_1$ ratio that features the QSL phase sandwiched between the 120$\degree$ N\'eel and stripe antiferromagnetic (AFM) states~\cite{Sherman165146,Iqbal144411}. Moreover, magnetic anisotropy which is also inherently present in real materials influences the ground state appreciably. TLAFs with easy-axis anisotropy favors double  magnetic transitions in zero-field where the collinear up-up-down ($uud$) phase becomes stable above the low temperature 120$\degree$ state~\cite{Harrison679,*Melchy064411,Miyashita3605,Lal014429,Ranjith014415}. On the other hand, a single step transition is predicted for dominant easy plane anisotropy which destabilizes the $uud$ phase and entails only 120$\degree$ state in zero-field~\cite{Miyashita3385,Rawl054412,Lee224402,Smirnov134412}. Few examples of the celebrated compounds in this category include Cs$_2$Cu$X_4$ ($X = $ Cl, Br) and $A_3$$T$$Z_2$O$_9$ ($A =$ Ba, Sr, Ca; $T = $Co, Ni, Mn; $Z = $ Ta, Nb, Sb)~\cite{Coldea1335,Ono104431,Susuki267201,lu094412,Lee104420,Lee224402}. These findings stimulate further interest in frustrated magnetism to look for new and exciting compounds experimentally.

\begin{figure}
	\includegraphics[width= \linewidth]{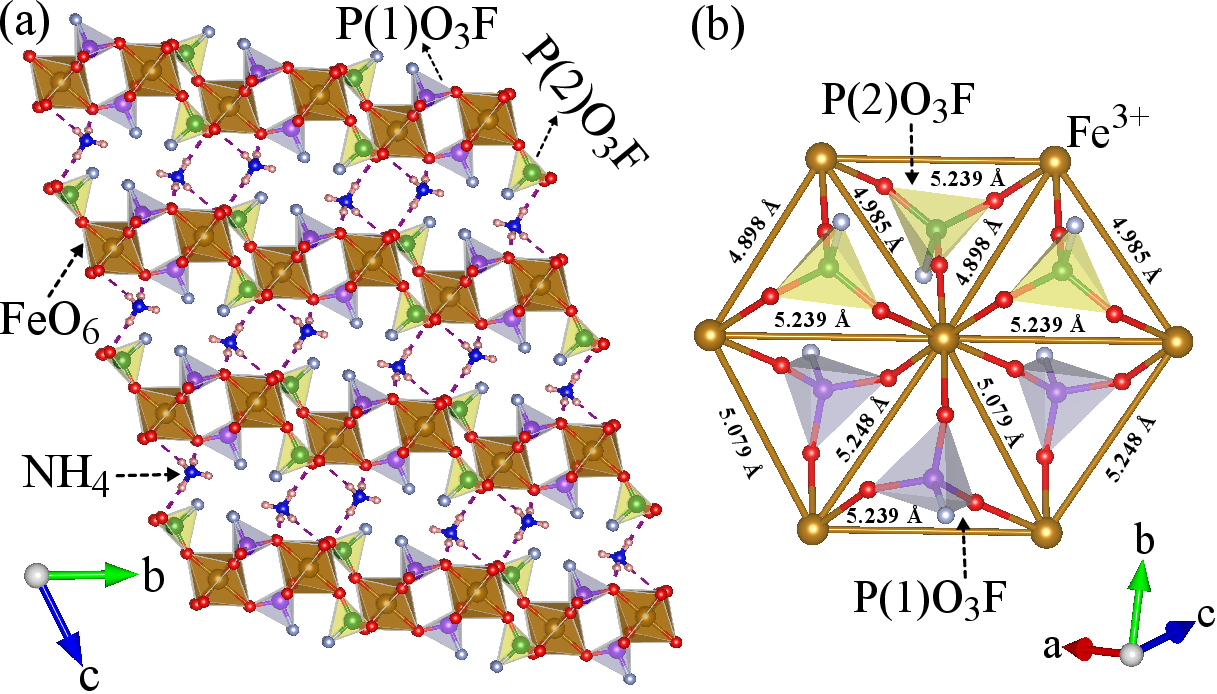}
	\caption{\label{Fig1}(a) Crystal structure of NH$_{4}$Fe(PO$_{3}$F)$_2$, showing adjacent triangular layers comprising of FeO$_6$ octahedra and PO$_3$F tetrahedra [P(1)O$_3$F and P(2)O$_3$F]. The hydrogen positions and bonds (dashed lines) in the NH$_4$ units between two layers are shown. (b) A section of the layer showing Fe$^{3+}$ ions connected via PO$_3$F tetrahedra on a triangular lattice. The Fe$^{3+}$ - Fe$^{3+}$ bond lengths in the triangular units are highlighted.}
\end{figure}
In this paper, we thoroughly studied the magnetic properties of a triangular lattice compound NH$_{4}$Fe(PO$_{3}$F)$_2$. The crystal structure of NH$_{4}$Fe(PO$_{3}$F)$_2$ is presented in Fig.~\ref{Fig1}(a). The distorted FeO$_6$ octahedra are corner shared with two inequivalent and distorted PO$_3$F tetrahedra [P(1)O$_3$F and P(2)O$_3$F] and form the triangular layers in the $ab$-plane. Two adjacent layers are well separated from each other via non-magnetic NH$_{4}^{+}$ ions. As the magnetic Fe$^{3+}$ ions in the adjacent layers are interconnected by weak hydrogen bonds, this may lead to a negligible interlayer coupling, making the system a frustrated 2D triangular lattice.
The 2D triangular layers are also slightly buckled. In addition, it was found that the Fe-Fe bond lengths in each triangular unit are unequal, resulting in a scalene triangle [see Fig.~\ref{Fig1}(b)]. Our magnetic measurements reveal that NH$_{4}$Fe(PO$_{3}$F)$_2$ is a moderately frustrated magnet and it undergoes a magnetic ordering at $T_{\rm N} \simeq 5.7$~K which is of commensurate AFM type. We observed three different phases in the $H-T$ phase diagram with a spin-flop transition, confirming magnetic anisotropy in the system.

\section{Experimental Details}
\begin{figure}
\includegraphics[width=\columnwidth]{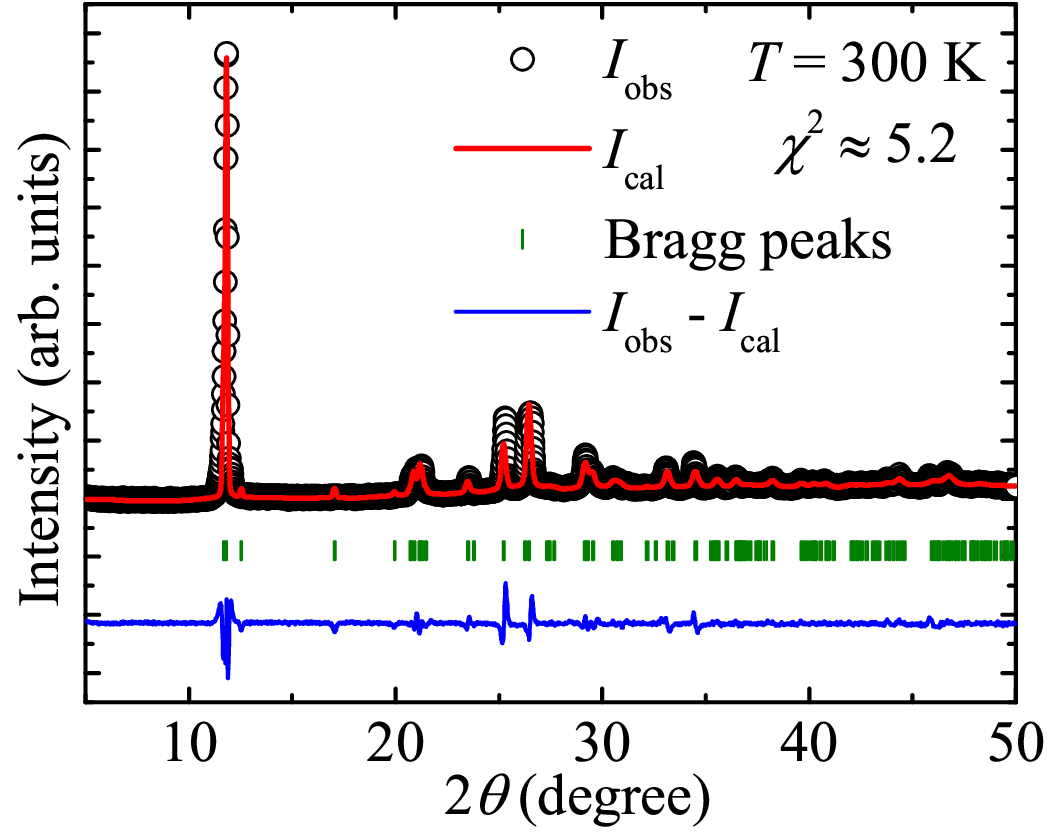}
\caption{\label{Fig2} 
Powder XRD data collected at room temperature. The red solid line is the Le Bail fit to the data, the vertical bars are the Bragg positions, and the blue line at the bottom is the difference between experimentally observed and calculated intensities.}
\end{figure}
Polycrystalline sample of NH$_{4}$Fe(PO$_{3}$F)$_2$ was synthesized by the conventional ionothermal reaction technique. 0.244~g FeCl$_3$ (Aldrich, 99.99\%), 0.230~g (NH$_{4}$)H$_{2}$PO$_{4}$ (Aldrich, 99.99\%), 0.108~g H$_{3}$BO$_{3}$ (Aldrich, 99.99\%), and 0.520~g [C$_{4}$mPy][PF$_{6}$] (Aldrich, 99.99\%) powders were taken in a 23~ml teflon lined bomb under an argon atmosphere, and heated at 160~$^0$C for 6 days followed by slow cooling (1.5~$^0$C/hour) to room temperature. The obtained solid product was filtered off and washed carefully with both deionized water and acetone in multiple rounds by centrifuge technique until the wash solution was colorless. The white solid product was dried in an oven at 80$^0$C for 12 hours and ground into fine powder for further characterizations. The phase purity of the product was confirmed by powder x-ray diffraction (XRD) recorded at room temperature using a PANalytical x-ray diffractometer (Cu\textit{K$_{\alpha}$} radiation, $\lambda_{\rm avg}\simeq 1.5418$ \AA). The XRD result was analyzed by Le Bail fit using the \texttt{FULLPROF} software package~\cite{Carvajal55} as shown in Fig.~\ref{Fig2}. With the help of Le Bail refinement, all the diffraction peaks of NH$_{4}$Fe(PO$_{3}$F)$_2$ could be indexed with the triclinic unit cell [$P\bar{1}$ (No.~2)], taking the initial structural parameters from Ref.~\cite{Stefanie7982}. The absence of any unidentified peaks suggest the phase purity of the polycrystalline sample. The obtained lattice parameters at room temperature are $a = 5.244(2)$~\AA, $b = 8.297(2)$~\AA, $c = 8.457(2)$~\AA, $\alpha = 64.271(1)$$^{\circ}$, $\beta = 89.695(1)$$^{\circ}$, $\gamma = 89.237(1)$$^{\circ}$ and the unit-cell volume $V_{\rm cell} = 332.28(1)$~\AA$^3$, which are in close agreement with the previous report~\cite{Stefanie7982}.

Magnetization ($M$) measurement was performed as a function of temperature (1.8~K~$\leq T \leq$~380~K) and magnetic field (0~$\leq H \leq$~7~T) using a superconducting quantum interference device (SQUID) (MPMS-3, Quantum Design) magnetometer. The high-field magnetization was measured at $T = 1.5$~K in pulsed magnetic field up to 55~T at the Dresden high magnetic field laboratory~\cite{Skourski214420,*Tsirlin132407}. Heat capacity ($C_{\rm p}$) as a function of $T$ (1.9~K~$\leq T \leq$~250~K) and $H$ (0~$\leq H \leq 9$~T) was measured on a small piece of sintered pellet using the thermal relaxation technique in the physical property measurement system (PPMS, Quantum Design). 

The nuclear magnetic resonance (NMR) measurements were carried out using spin-echo method on the $^{31}$P nuclei (nuclear spin $I=1/2$ and gyromagnetic ratio $\gamma_{N}/2\pi = 17.237$~MHz/T). Since $^{31}$P does not involve electric quadrupole effects, $^{31}$P NMR is an ideal probe to study magnetism and spin fluctuations. We performed the experiments at two magnetic fields ($\mu_{\rm 0}H = 1.3$ and 7~T) and over a wide temperature range (1.5~K $\leq T \leq$ 300~K). The $^{31}$P NMR spectra were obtained by sweeping the radio frequency, keeping the magnetic field fixed. The temperature-dependent NMR shift, $K(T)=[\nu-\nu_{\rm ref}]/\nu_{\rm ref}$ was calculated by taking the resonance frequency of the sample ($\nu$) with respect to the resonance frequency of a non-magnetic reference ($\nu_{\rm ref}$). $^{31}$P spin-lattice relaxation rate ($1/T_1$) was measured by the standard saturation recovery method. $^{31}$P spin-spin relaxation rate ($1/T_2$) was obtained by measuring the decay of the echo integral with variable spacing between the $\pi$/2 and $\pi$ pulses.

Full diagonalization (FD) was performed to simulate magnetic susceptibility using the \texttt{loop} and \texttt{fulldiag} algorithms~\cite{Todo047203} of the \texttt{ALPS} package~\cite{BauerP05001}. 

\section{Results and Discussion}
\subsection{Magnetization}
\begin{figure}
	\includegraphics[width=\linewidth]{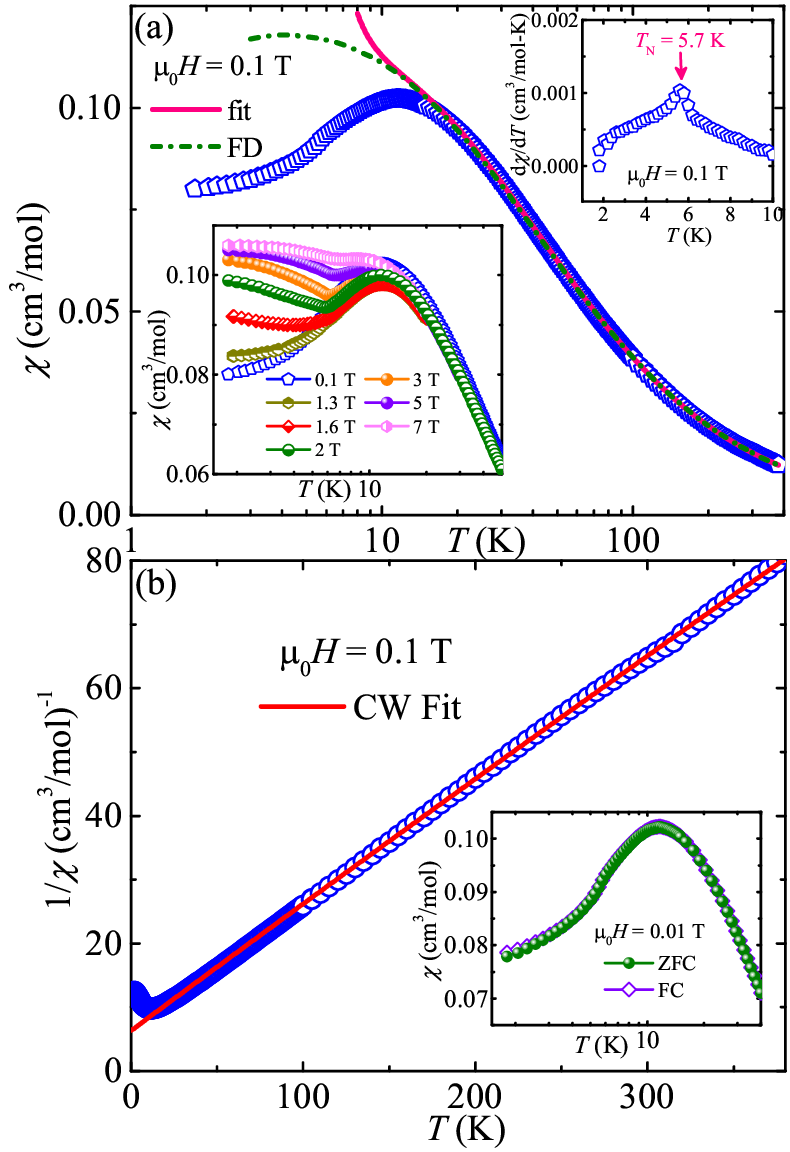}
	\caption{(a) Magnetic susceptibility $\chi(T)$ measured in an applied field of $\mu_{\rm 0}H = 0.1$~T. The solid line represents the fit using isotropic triangular lattice model [Eq.~\eqref{cw1}]. The dash-dotted line is the simulation of $\chi(T)$ of a spin-$5/2$ isotropic triangular lattice using full diagonalization (FD). Upper inset: $d\chi/dT$ vs $T$ measured in $\mu_{\rm 0}H = 0.1$~T highlighting the transition temperature ($T_{\rm N}$). Lower inset: $\chi(T)$ measured in different fields. (b) $1/\chi$ vs $T$ at $\mu_{\rm 0}H = 0.1$~T and the solid line is the CW fit. Inset: $\chi(T)$ measured at $H = 100$~Oe in ZFC and FC protocols.}
	\label{Fig3}
\end{figure}
\begin{figure}
	\includegraphics[width=\linewidth]{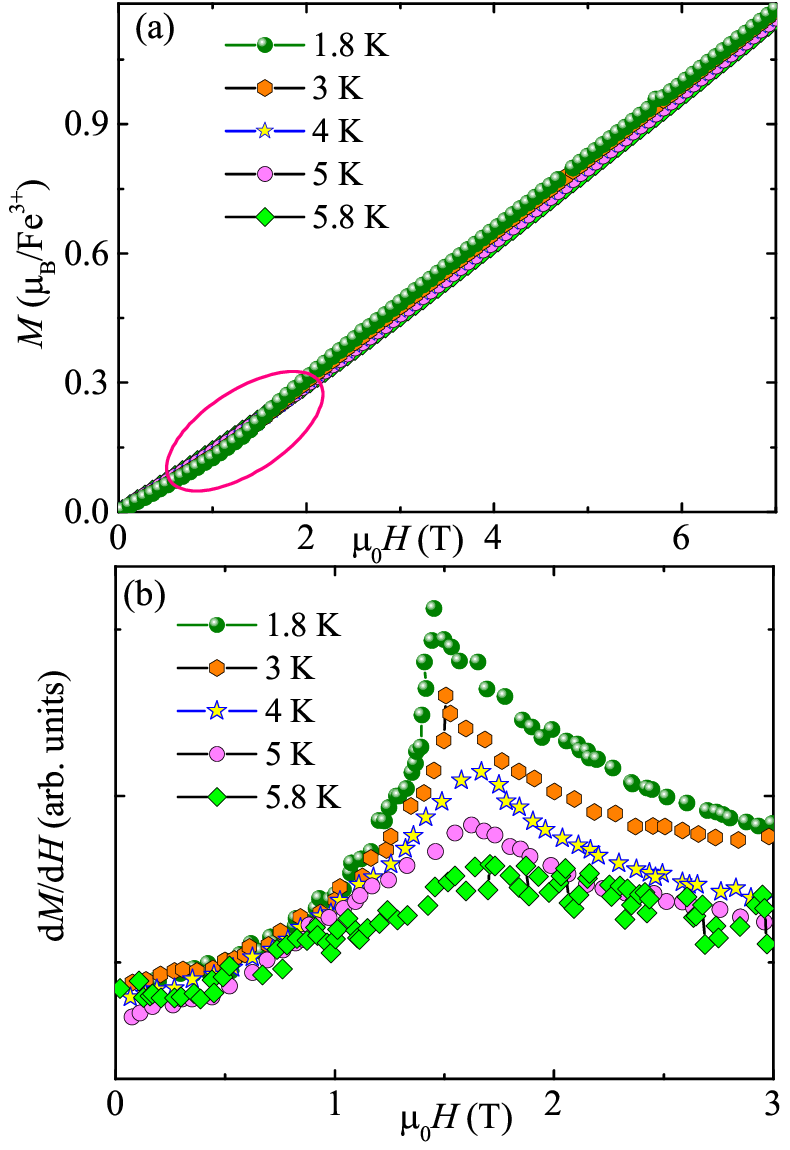}
	\caption{(a) Magnetic isotherms ($M$ vs $H$) measured at various temperatures from $T=1.8$ to 5.8~K. The encircled region marks as the field induced transition regime. (b) Derivative of isothermal magnetization vs $H$ in different temperatures highlighting the field induced transition.}
	\label{Fig4}
\end{figure}
Temperature dependent dc susceptibility $\chi(T)$ ($\equiv M/H$) of the polycrystalline NH$_{4}$Fe(PO$_{3}$F)$_2$ sample measured in an applied field of $\mu_{\rm 0}H = 0.1$~T is shown in Fig.~\ref{Fig3}(a). At high temperatures, $\chi(T)$ increases with decreasing temperature in a Curie-Weiss (CW) manner and passes through a broad maximum at $T^{\rm{max}}_{\chi} \simeq 12$~K indicative of an AFM short-range order, typically expected in low-dimensional quantum magnets. A kink is observed at $T_{\rm{N}}\simeq 5.7$~K, indicating the transition to a three-dimensional (3D) magnetic LRO state possibly triggered by a weak interlayer coupling. This transition is more pronounced in the d$\chi$/d$T$ vs $T$ plot shown in the upper inset of Fig.~\ref{Fig3}(a). To understand the nature of the magnetic transition, $\chi(T)$ was measured at different magnetic fields up to 7~T, as shown in the lower inset of Fig.~\ref{Fig3}(a). With increasing $H$, $T_{\rm N}$ remains unchanged up to 1.6~T and then moves towards high temperatures above 1.6~T. For $H \geq 1.6$~T, $\chi(T)$ below $T_{\rm N}$ develops a contort, which emphasizes that there is some kind of field-induced spin canting prevailing in the system~\cite{Garlea011038}.

Figure~\ref{Fig3}(b) presents the inverse susceptibility [$1/\chi(T)$] for $\mu_{\rm 0}H = 0.1$~T. In the high temperature paramagnetic regime, $1/\chi(T)$ typically shows a linear behavior with temperature, due to uncorrelated moments. To extract the magnetic parameters, $1/\chi(T)$ was fitted in the temperature range 135~K~$\leq T \leq 380$~K by the Curie-Weiss (CW) law
\begin{equation}
	\chi(T)=\chi_0+\frac{C}{T-\theta_{\rm CW}},
	\label{cw}
\end{equation}
where, $\chi_0$ is the temperature-independent susceptibility that includes Van-Vleck paramaganetism and core diamagnetism. The second term is the CW law where $C$ is the Curie constant and $\theta_{\rm CW}$ is the CW temperature. The fit yields $\chi_{0} \simeq 3.75 \times 10^{-4}$~cm$^{3}$/mol, $C \simeq 4.96$~cm$^{3}$ K/mol, and $\theta_{\rm CW} \simeq -33$~K. The large negative value of $\theta_{\rm CW}$ suggests that the dominant exchange interaction between Fe$^{3+}$ ions is AFM in nature. From the value of $C$, the effective moment is calculated to be $\mu_{\rm eff} \simeq 6.30\mu_{\rm B}$ using the relation $\mu_{\rm eff} = \sqrt{(3k_{\rm B}C/N_{\rm A})}\mu_{\rm B}$, where $N_{\rm A}$ is the Avogadro's number and $\mu_{\rm B}$ is the Bohr magneton. For a spin-$5/2$ system (Fe$^{3+}$), the spin-only effective moment is expected to be $\mu_{\rm eff} = g\sqrt{S(S+1)} \mu_{\rm B} \simeq 5.91\mu_{\rm B}$, assuming Land$\acute{e}$ $g$-factor $g\simeq 2$. Our experimental value of $\mu_{\rm eff} \simeq 6.30\mu_{\rm B}$ is slightly higher than the spin-only value and corresponds to $g\simeq 2.12$. The core diamagnetic susceptibility $\chi_{\rm core}$ of NH$_{4}$Fe(PO$_{3}$F)$_2$ is calculated to be $-1.15 \times 10^{-4}$~cm$^{3}$/mol by adding the core diamagnetic susceptibilities of individual ions NH$_{4}^{+}$, Fe$^{3+}$,  P$^{5+}$, O$^{2-}$, and F$^{1-}$~\cite{Selwood1956,*Bain532}. The Van-Vleck paramagnetic susceptibility ($\chi_{\rm VV}$) is estimated by subtracting $\chi_{\rm dia}$ from $\chi_{0}$ to be $\sim 4.9 \times 10^{-4}$~cm$^{3}$/mol. Inset of Fig.~\ref{Fig3}(b) displays the zero-field-cooled (ZFC) and field-cooled (FC) susceptibilities in $\mu_{\rm 0}H = 0.01$~T. The absence of splitting between ZFC and FC data rules out the possibility of any spin-glass transition or spin-freezing at low temperatures.

As mentioned earlier, magnetic frustration impels the magnetic LRO towards lower temperatures compared to the AFM exchange energy. Further, since $\theta_{\rm CW}$ represents the overall energy scale of the exchange couplings, the relatively reduced value of $T_{\rm N}$ with respect to $\theta_{\rm CW}$ is considered as an empirical measure of the effect of frustration. In NH$_{4}$Fe(PO$_{3}$F)$_2$, the frustration ratio is calculated to be $f = \frac{|\theta_{\rm CW}|}{T_{\rm N}} \simeq 5.7$, which corroborates moderate magnetic frustration in the system. From the value of $\theta_{\rm CW}$, one can also estimate the average exchange coupling as $|\theta_{\rm CW}| = \frac{JzS(S+1)}{3 k_{\rm B}}$, assuming a isotropic triangular lattice antiferromagnet~\cite{Domb296}. Here, $z = 6$ is the number of nearest neighbours of Fe$^{3+}$ ions. Using the experimental value of $\theta_{\rm CW}$, the average exchange coupling within the triangular planes is estimated to be $J/k_{\rm B} \simeq 1.89$~K.

To estimate the exchange coupling between Fe$^{3+}$ ions, $\chi(T)$ is decomposed into two components
\begin{equation}
	\chi(T)=\chi_0+\chi_{\rm spin}(T).
	\label{cw1}
\end{equation}
Here, $\chi_{\rm spin}(T)$ is the spin susceptibility which can be chosen according to the intrinsic magnetic model. As the Fe$^{3+}$ ions in the crystal structure are arranged in triangles, we took the expression of the high temperature series expansion (HTSE) of $\chi_{\rm spin}$ for a spin-5/2 Heisenberg isotropic TLAF model which has the form~\cite{Delma55,Sebastian104425}
\begin{equation}
	\label{TLA}
	\frac {N_{\rm A} g^2 \mu_{\rm B}^2} {3|J|\chi_{\rm spin}}= x+4+\frac {3.20}{x}-\frac {2.186}{x^2}+\frac {0.08}{x^3}+\frac {3.45}{x^4}-\frac {3.99}{x^5}.
\end{equation}
Here, $x = k_{\rm B}T/{|J|S(S+1)}$. This expression is valid for $T \geq JS(S+1)$~\cite{Schmidt104443}. The solid line in Fig.~\ref{Fig3}(a) represents the best fit to the $\chi(T)$ data above 18~K by the Eq.~\eqref{TLA} resulting in $\chi_{0} \simeq 1.50 \times 10^{-4}$~cm$^{3}$/mol, $g\simeq 2.12$, and the average AFM exchange coupling $J/k_{\rm B} \simeq 1.7$~K. This value is in good agreement with the value estimated from the $\theta_{\rm CW}$ analysis.

Figure~\ref{Fig4}(a) presents the magnetic isotherms ($M$ vs $H$) measured at various temperatures up to 7~T. For $T\leq 5.8$~K a bend is observed in the intermediate magnetic fields which is the signature of a meta-stable field-induced transition. This bend is more pronounced in the d$M$/d$H$ vs $H$ plots shown in Fig.~\ref{Fig4}(b), where this feature is manifested as a well defined peak. For $T=1.8$~K, d$M$/d$H$ shows the peak at 1.45~T. As the temperature rises, the peak moves weakly towards higher fields and disappears completely for $T > 5.8$~K.

\begin{figure}
	\includegraphics[width=\linewidth]{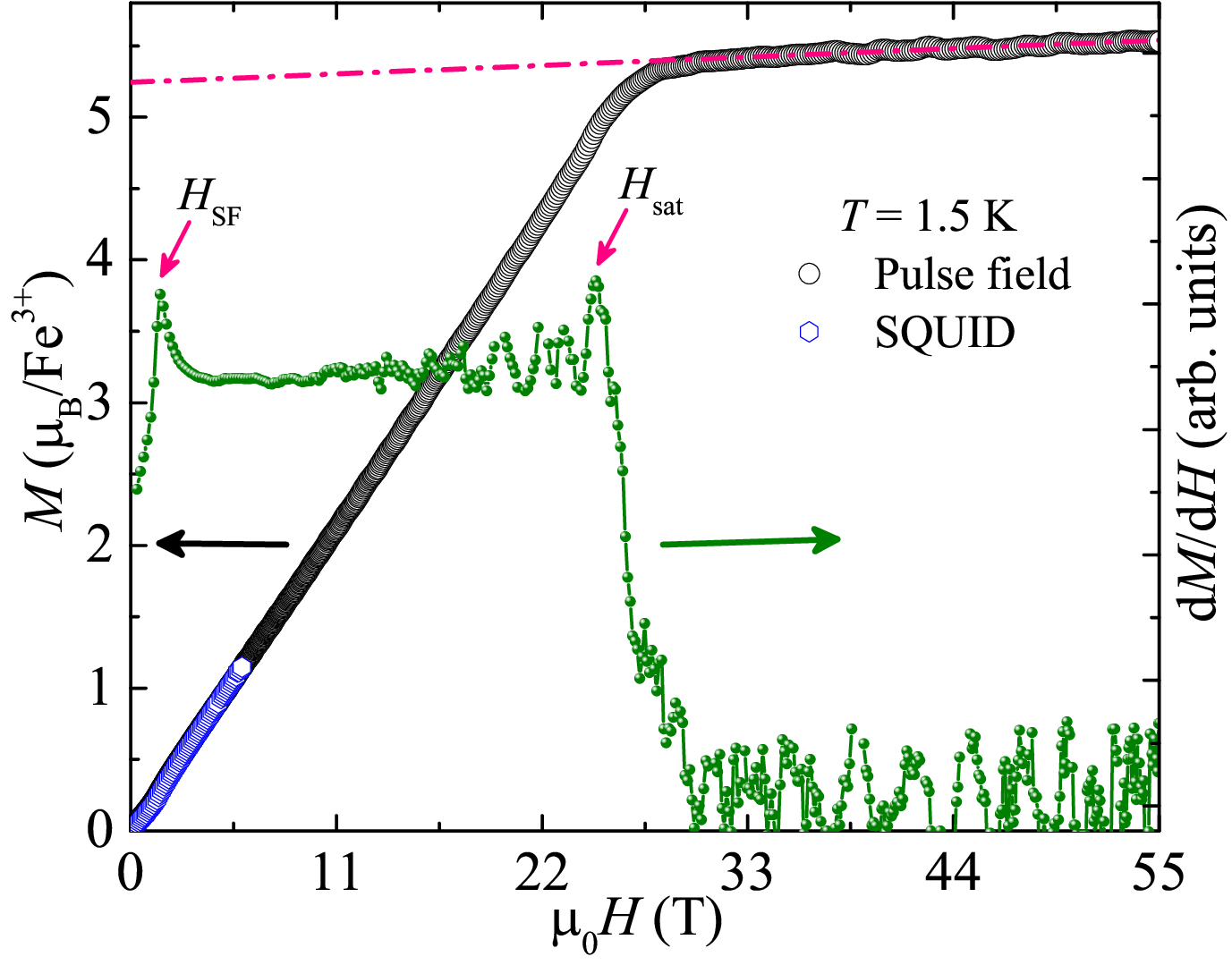}
	\caption{Magnetization ($M$) and its derivative ($dM/dH$) vs magnetic field ($H$) at $T=1.5$~K measured using pulsed magnetic field in the left and right $y$-axes, respectively. The pulse field data are scaled with respect to the SQUID data. The dash-dotted line represents the saturation magnetization. The sharp peaks in $dM/dH$ vs $H$ plot (highlighted by arrows) depict the spin-flop ($H_{\rm SF}$) and saturation ($H_{\rm sat}$) fields, respectively.}
	\label{Fig5}
\end{figure}
Since no saturation in magnetization was achieved up to 7~T, we measured $M$ vs $H$ up to 55~T using pulse field at $T = 1.5$~K (Fig.~\ref{Fig5}). The pulse field data are quantified by scaling with respect to the SQUID data measured up to 7~T at $T=1.8$~K. $M$ increases almost linearly with $H$ and then exhibits a kink at $\mu_{\rm 0}H_{\rm SF} = 1.45$~T, similar to that observed in Fig.~\ref{Fig4}(b), reminiscent of a spin-flop (SF) transition. Further increase in $H$, $M$ increases linearly and develops a sharp bend towards saturation above $\mu_{\rm 0}H_{\rm sat} \simeq 25$~T. These two critical fields are very well visualized as sharp peaks in the d$M$/d$H$ vs $H$ plot. No obvious feature associated with the $1/3$ magnetization plateau is observed in the intermediate field range. In the polycrystalline sample, the random orientation of the grains with respect to the applied field direction might obscure this plateau. Above $H_{\rm sat}$, $M$ shows a slightly upward trend which may be due to a small Van-Vleck paramagnetic contribution. The linear interpolation of a straight line fit to the magnetization above $H_{\rm sat}$ intercepts the $y$-axis at $M_{\rm sat}\simeq 5.24\mu_{\rm B}$. This saturation magnetization corresponds to the value expected for a spin-5/2 ion with $g = 2.1$. Further, in an antiferromagnetically ordered spin system $H_{\rm sat}$ defines the energy required to overcome the AFM interactions and to polarize the spins in the direction of applied field. In particular, in a Heisenberg TLAF, $H_{\rm sat}$ can be written in terms of the intralayer exchange coupling as, $H_{\rm sat} = \frac{9JS}{g \mu_{\rm B}}$~\cite{Sebastian104425}. Our experimental value of $\mu_{\rm 0}H_{\rm sat} \simeq 25$~T yields an average exchange coupling of $J/k_{\rm B} \simeq 1.73$~K, which is indeed close to the value obtained from the analysis of $\chi(T)$ and $\theta_{\rm CW}$. A small difference can be attributed to the magnetic anisotropy present in the compound.

\subsection{Heat Capacity}
\begin{figure}
	\includegraphics[width=\linewidth]{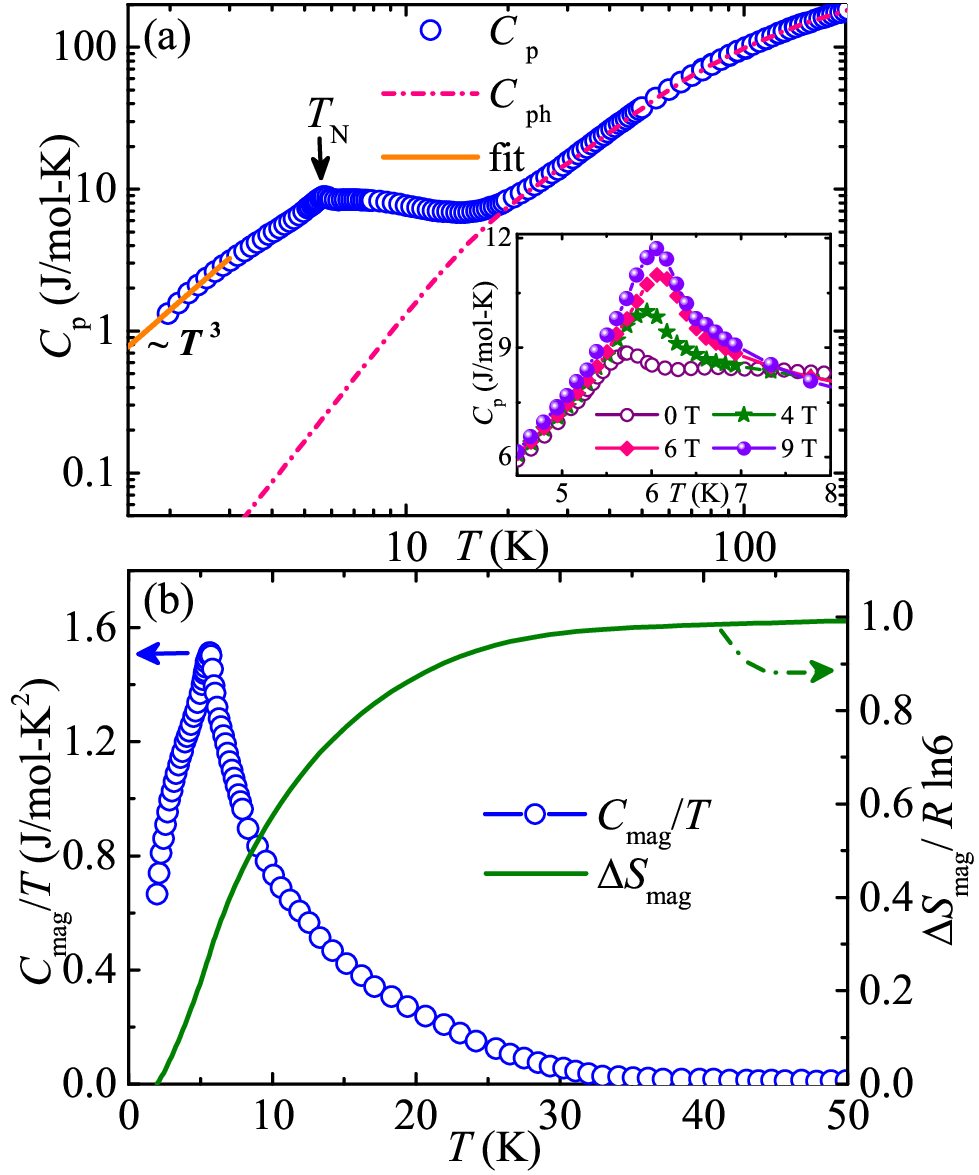}
	\caption{(a) $C_{\rm p}$ vs $T$ of NH$_{4}$Fe(PO$_{3}$F)$_2$ measured in zero applied field along with the low temperature power law fit (solid line). The dash-dotted line represents the calculated phononic contribution ($C_{\rm ph}$). Inset: Field dependence of $C_{\rm p}$($T$) around $T_{\rm N}$. (b) $C_{\rm mag}/T$ and ${\Delta}S_{\rm mag}$ vs $T$ in the left and right $y$-axes, respectively.}
	\label{Fig6}
\end{figure}
Temperature dependent heat capacity ($C_{\rm p}$) measured in zero magnetic field is presented in Fig.~\ref{Fig6}(a). As the temperature is lowered, $C_{\rm p}$ goes down systematically and then shows a pronounced $\lambda$-type anomaly at $T_{\rm N}\simeq 5.7$~K, confirming the onset of magnetic LRO. Well below $T_{\rm N}$, $C_{\rm p}(T)$ follows a $T^3$ behaviour [solid line in Fig.~\ref{Fig6}(a)], portraying the dominance of three-dimensional (3D) magnon excitations~\cite{Somesh104422}. In a magnetic insulator, the total heat capacity $C_{\rm p}(T)$ is the sum of two major contributions: phonon contribution $C_{\rm ph}(T)$ which dominates in the high-temperature region and magnetic contribution $C_{\rm mag}(T)$ that dominates in the low-temperature region. 
In order to extract $C_{\rm mag}(T)$ from $C_{\rm p}(T)$, the phonon contribution $C_{\rm ph}(T)$ was first estimated by fitting the high-$T$ data by a linear combination of one Debye [$C_{\rm D}(T)$] and three Einstein [$C_{\rm E}(T)$] terms (Debye-Einstein model) as~\cite{Mohanty134401,*Gopal2012}
\begin{equation}
	C_{\rm ph}(T)=f_{\rm D}C_{\rm D}(\theta_{\rm D},T)+\sum_{i = 1}^{3}g_{i}C_{{\rm E}_i}(\theta_{{\rm E}_i},T).
	\label{Eq5}
\end{equation}
The first term in Eq.~\eqref{Eq5} takes into account the acoustic modes, called the Debye term with the coefficient $f_{\rm D}$ and 
\begin{equation}
	C_{\rm D} (\theta_{\rm D}, T)=9nR\left(\frac{T}{\theta_{\rm D}}\right)^{3} \int_0^{\frac{\theta_{\rm D}}{T}}\frac{x^4e^x}{(e^x-1)^2} dx.
	\label{Eq6}
\end{equation}
Here, $x=\frac{\hbar\omega}{k_{\rm B}T}$, $\omega$ is the frequency of oscillation, $R$ is the universal gas constant, and $\theta_{\rm D}$ is the characteristic Debye temperature.
The second term in Eq.~\eqref{Eq5} accounts for the optical modes of the phonon vibration, known as the Einstein term with the coefficient $g_i$ and
\begin{equation}
C_{\rm E}(\theta_{\rm E}, T) = 3nR\left(\frac{\theta_{\rm E}}{T}\right)^2 
\frac{e^{\left(\frac{\theta_{\rm E}}{T}\right)}}{[e^{\left(\frac{\theta_{\rm E}}{T}\right)}-1]^{2}}.
\label{Eq7} 
\end{equation}
Here, $\theta_{\rm E}$ is the characteristic Einstein temperature. The coefficients $f_{\rm D}$, $g_1$, $g_2$, and $g_3$ represent the fraction of atoms that contribute to their respective parts. These values are taken in such a way that their sum should be equal to one.

The zero field $C_{\rm p}(T)$ data above $\sim 30$~K are fitted by Eq.~\eqref{Eq5} [dotted line in Fig.~\ref{Fig5}(a)] and the obtained parameters are $f_{\rm D} \simeq 0.062$, $g_1 \simeq 0.242$, $g_2 \simeq 0.3$, $g_3 \simeq 0.395$, $\theta_{\rm D} \simeq 113$~K, $\theta_{{\rm E}_1} \simeq 245$~K, $\theta_{{\rm E}_2} \simeq 550$~K, and $\theta_{{\rm E}_3} \simeq 1500$~K. Finally, the high-$T$ fit was extrapolated down to low temperatures and $C_{\rm mag}(T)$ was estimated by subtracting $C_{\rm {ph}}(T)$ from $C_{\rm p}(T)$. Figure~\ref{Fig5}(b) presents $C_{\rm mag}(T)/T$ and the corresponding magnetic entropy [$S_{\rm mag}(T) = \int_{\rm 1.9\,K}^{T}\frac{C_{\rm {mag}}(T')}{T'}dT'$] in the left and right $y$-axes, respectively. The obtained magnetic entropy reaches a maximum value of $S_{\rm mag} \simeq 14.83$~J/mol-K around 30~K and this value is close to the expected theoretical value of $S_{\rm mag} = R \ln(2S+1)=14.89$~J/mol-K for a $S = 5/2$ system.

To gain more information about the magnetic transition, we measured $C_{\rm p}(T)$ in different applied fields [inset of Fig.~\ref{Fig6}(a)]. With increasing magnetic field upto $\mu_{\rm 0}H = 6$~T, the height of the peak is enhanced substantially and the peak position shifts towards high temperatures, consistent with the $\chi(T)$ data. With further increase in field above 6~T, $T_{\rm N}$ shifts towards low-temperatures.

\subsection{$^{31}$P NMR}
NMR is an effective local probe for examining the static and dynamic characteristics of a spin system. The crystal structure of NH$_{4}$Fe(PO$_{3}$F)$_2$ has two inequivalent $^{31}$P sites [P(1) and P(2)] with equal occupancies and both of them are coupled to the Fe$^{3+}$ ions (see Fig.~\ref{Fig1}). Both the $^{31}$P sites reside in an almost symmetric position between the Fe$^{3+}$ ions. As $^{31}$P has nuclear spin $I = 1/2$, one could expect a narrow and single spectral line for each P site.

\subsubsection{$^{31}$P NMR Spectra ($T > T_{\rm N}$)}
\begin{figure}
	\includegraphics[width=0.48\textwidth]{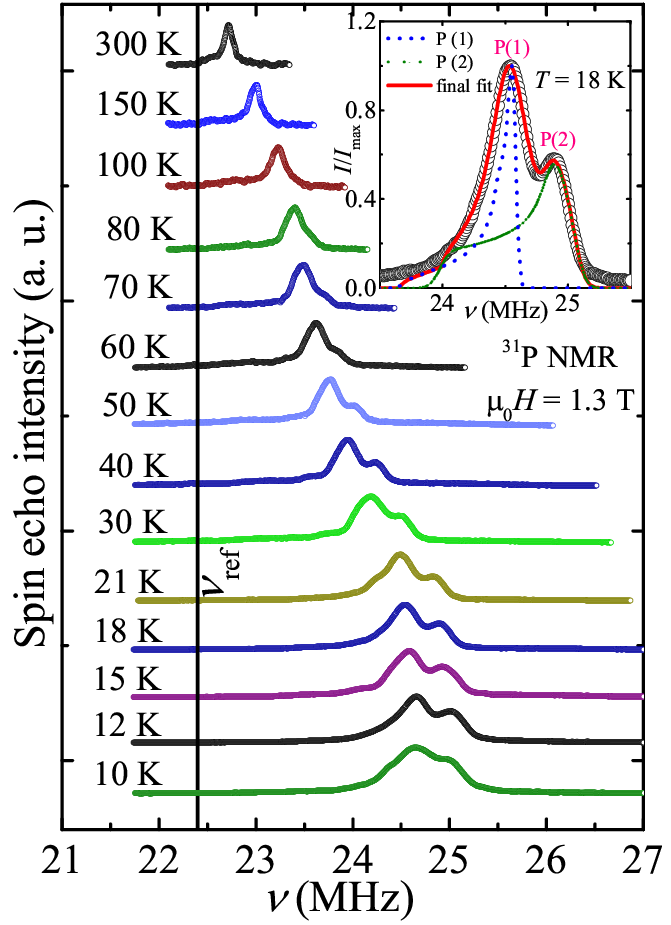}
	\caption{Temperature evolution of $^{31}$P NMR spectra measured in a magnetic field of $\mu_{\rm 0}H = 1.3$~T down to $T = 10$~K (above $T_{\rm N}$). The vertical solid line corresponds to the $^{31}$P non-magnetic reference frequency position. Inset: Spectral fit for $T = 18$~K taking the superposition of two inequivalent P sites [P(1) and P(2)].}
	\label{Fig7}
\end{figure}
The frequency-sweep $^{31}$P NMR spectra were measured at different temperatures and in two different magnetic fields (1.3 and 7~T). Note that 1.3~T and 7~T are well below and above the SF critical field $H_{\rm SF}$, respectively. Figure~\ref{Fig7} presents the $^{31}$P NMR spectra at 1.3~T measured down to 10~K (above $T_{\rm N}$), where, each spectrum is normalized by its maximum amplitude and vertically offset by adding a constant. At high temperatures (down to 100~K), a single and nearly symmetric spectral line is observed, typical for a $I=1/2$ nucleus~\cite{Mohanty104424,*Baek134424}. A single spectral line also implies that both the P-sites experience almost the same local environment at high temperatures. The spectral line becomes asymmetric with decrease in temperature. This line shape can be attributed to either the presence of two P sites in the crystal structure or a powder pattern due to an asymmetric hyperfine coupling constant and/or anisotropic susceptibility~\cite{Yogi024413}.
Below about 100~K, the line broadens with a shoulder on the high-frequency side. This shoulder becomes distinct and transforms into an additional peak for $T < 60$~K. The main peak is assigned as P(1) site while the sister peak is assigned as P(2) site.
Both the peaks shift towards high frequency side as we lower the temperature. This suggests that the two P sites experience different hyperfine fields at low temperatures.

\begin{figure}
	\includegraphics[width=\linewidth]{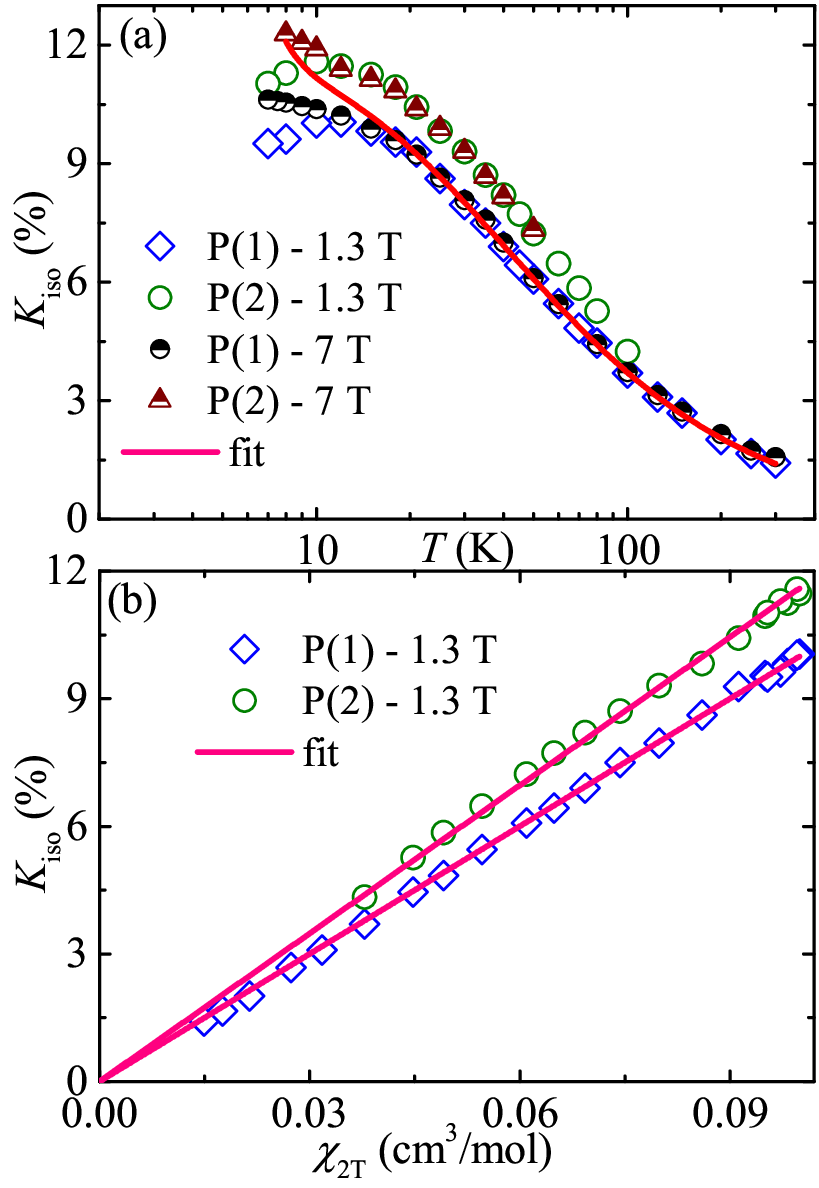}
	\caption{(a) Temperature dependent $^{31}$P NMR iso-shift measured in magnetic fields of $\mu_{\rm 0}H = 1.3$ and 7~T for both P(1) and P(2) sites. The solid line corresponds to the isotropic triangular lattice model fit using Eq.~\eqref{K}. (b) $K_{\rm iso}$ vs $\chi$ (measured in 2~T) for both P(1) and P(2) sites. The solid lines are straight line fits.}
	\label{Fig8}
\end{figure}
To evaluate the NMR iso-shift [$K_{\rm iso}$], each spectrum above 100~K is fitted by a single Gaussian function while below 100~K, it is fitted as the superposition of two asymmetric line shapes corresponding to two P sites. In the inset of Fig.~\ref{Fig7}, we have shown the spectral fit at $T = 18$~K considering two asymmetric lines. The area under both the peaks is found to be nearly equal.
A small discrepancy between the simulated and the experimental spectra can be attributed to the effect of partial orientation of grains with respect to the external field and/or anisotropic spin-spin relaxation time. The NMR iso-shift obtained by fitting the NMR spectra for 1.3 and 7~T are plotted as a function of temperature in Fig.~\ref{Fig8}(a). With decrease in temperature, $K_{\rm iso}(T)$ of both the P sites increases in a CW manner and then passes through a broad maxima around 12~K, similar to $\chi(T)$. Please note that $K_{\rm iso}(T)$ is a direct measure of intrinsic spin susceptibility $\chi_{\rm spin}(T)$ and is completely free from extrinsic contributions. Therefore, one can write $K_{\rm iso}(T)$ in terms of $\chi_{\rm spin}(T)$ as
\begin{equation}
	\label{K}
	K_{\rm iso}(T) = K_0 +\frac{A_{\rm hf }}{N_{\rm A}\mu_{\rm B}}\chi_{\rm spin}(T),
\end{equation}
where, $K_0$ is the temperature-independent chemical shift and $A_{\rm hf}$ is the hyperfine coupling constant between the $^{31}$P nuclei and the Fe$^{3+}$ electronic spins. In Fig.~\ref{Fig8}(b) we plotted $K_{\rm iso}$ vs $\chi$ with $T$ as an implicit parameter for 1.3~T for both the P-sites. The plots are linear over the whole temperature range (12~K to 300~K) and a straight line fit yields the hyperfine coupling constant $A_{\rm hf}^{\rm P(1)} \simeq 0.56$~T/$\mu_{\rm B}$ and $A_{\rm hf}^{\rm P(2)} \simeq 0.62$~T/$\mu_{\rm B}$ for P(1) and P(2) sites, respectively. Thus, both the P-sites are coupled to Fe$^{3+}$ ions with almost equal strength. These values are also comparable with the $A_{\rm hf}$ values reported for other transition metal phosphates~\cite{Islam174432,Devi012803}.

To estimate the exchange coupling between the Fe$^{3+}$ ions, $K(T)$ of P(1) site for 1.3~T is fitted by the Eq.~\eqref{K}, where $\chi_{\rm spin}$ is Eq.~\eqref{TLA}. During fitting, we fixed $A_{\rm hf}^{\rm{P(1)}} \simeq 0.56$~T/$\mu_{\rm B}$. As shown in Fig.~\ref{Fig8}(a), a good fit (solid line) above 15~K returns $J/k_{\rm B} \simeq 1.7$~K and $g \simeq 2.12$. This value of exchange coupling closely matches with the one obtained from the $\chi(T)$ analysis.

\subsubsection{$^{31}$P NMR Spectra ($T < T_{\rm N}$)}
\begin{figure}
\includegraphics[width=0.48\textwidth]{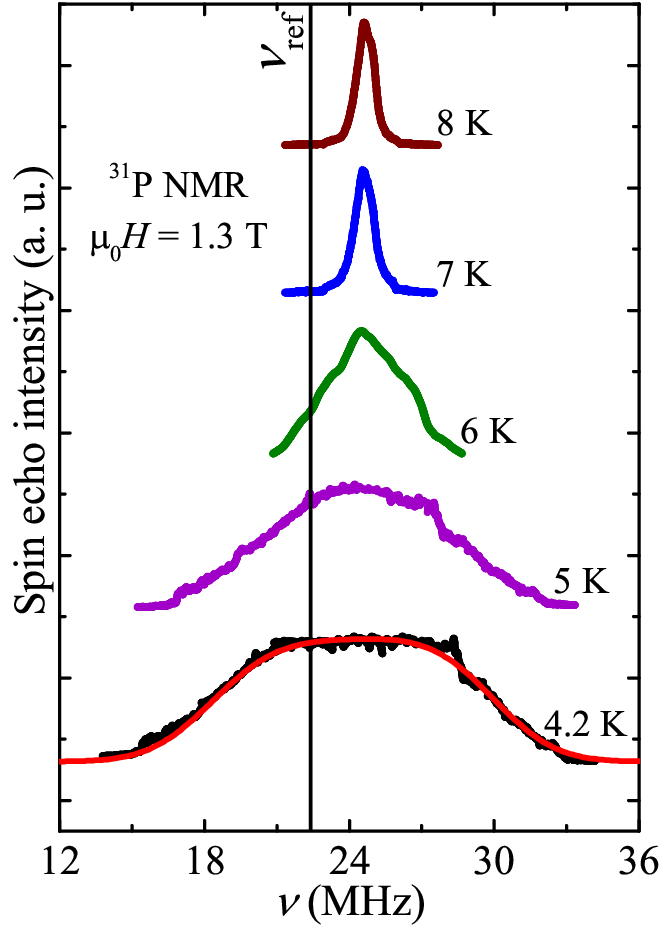}
\caption{$^{31}$P NMR spectra measured below 8~K for $\mu_0 H = 1.3$~T. The vertical solid line corresponds to $^{31}$P non-magnetic reference frequency (zero shift) position. The solid line is the fit of the spectrum at $T = 4.2$~K using Eq.~\eqref{convolv}.}
\label{Fig9}
\end{figure}
For $T < 8$~K, the line broadens drastically and both the peaks overlap on each other, resulting in a single spectral line. Such an abrupt line broadening implies the development of static internal field sensed by $^{31}$P, as one approaches the magnetic ordered state. As shown in Fig.~\ref{Fig9}, below $T_{\rm N} \simeq 6$~K, a huge internal field emerges leading to a drastic line broadening and the line attains a nearly rectangular shape. This rectangular line shape is reminiscent of a typical commensurate AFM ordering, which is due to the random distribution of the direction of the internal field with respect to the applied field $H$ in the powder sample~\cite{Ranjith014415,Ranjith024422,Yamada1751,Kikuchi2660}.

Assuming a uniform internal field in the ordered state, the NMR spectrum $f(H)$ can be expressed as~\cite{Yamada1751,Kikuchi2660,Nath024431}:
\begin{equation}
f(H) \propto \frac{H^2 - H_{\rm n}^2 + \omega_{\rm N}^2/\gamma_{\rm N}^2}{H_{\rm n}H^2},
\end{equation}
where $\omega_{\rm N}$ is the NMR frequency which is assumed to be larger than $\gamma_{\rm N}H_{\rm n}$ and $H_{\rm n}$ is the antiferromagnetic internal field. Two cutoff fields, $\frac{\omega_{\rm N}}{\gamma_{\rm N}} + H_n$ and $\frac{\omega_{\rm N}}{\gamma_{\rm N}} - H_n$ will produce two sharp edges of the spectrum. However, in real materials, these sharp edges are usually smeared by the inhomogeneous distribution of the internal field. In order to take this effect into account, we convoluted $f(H)$ with a Gaussian distribution function $g(H)$:
\begin{equation}
 F(H) = \int f(H - H')g(H') dH',
 \label{convolv}
\end{equation}
 where the distribution function is taken as~\cite{Kikuchi2660}
\begin{equation}
  g(H) = \frac{1}{\sqrt{2\pi\Delta H_{n}^2}} \exp\left(-\frac{(H - H_{n})^2}{2 \Delta H_n}\right).
\end{equation}
As shown in Fig.~\ref{Fig9}, the simulated curve using Eq.~\eqref{convolv} reproduces the experimental line shape very well at $T = 4.2$~K, confirming the commensurate nature of the ordered state below $T_{\rm N}\simeq 6$~K. Indeed, such a rectangular line shape has been observed in several compounds in the commensurate AFM ordered state~\cite{Yogi024413,Ranjith014415,Devi012803}.


\subsubsection{Spin-lattice relaxation rate $1/T_1$}
\begin{figure}
	\includegraphics[width=\linewidth]{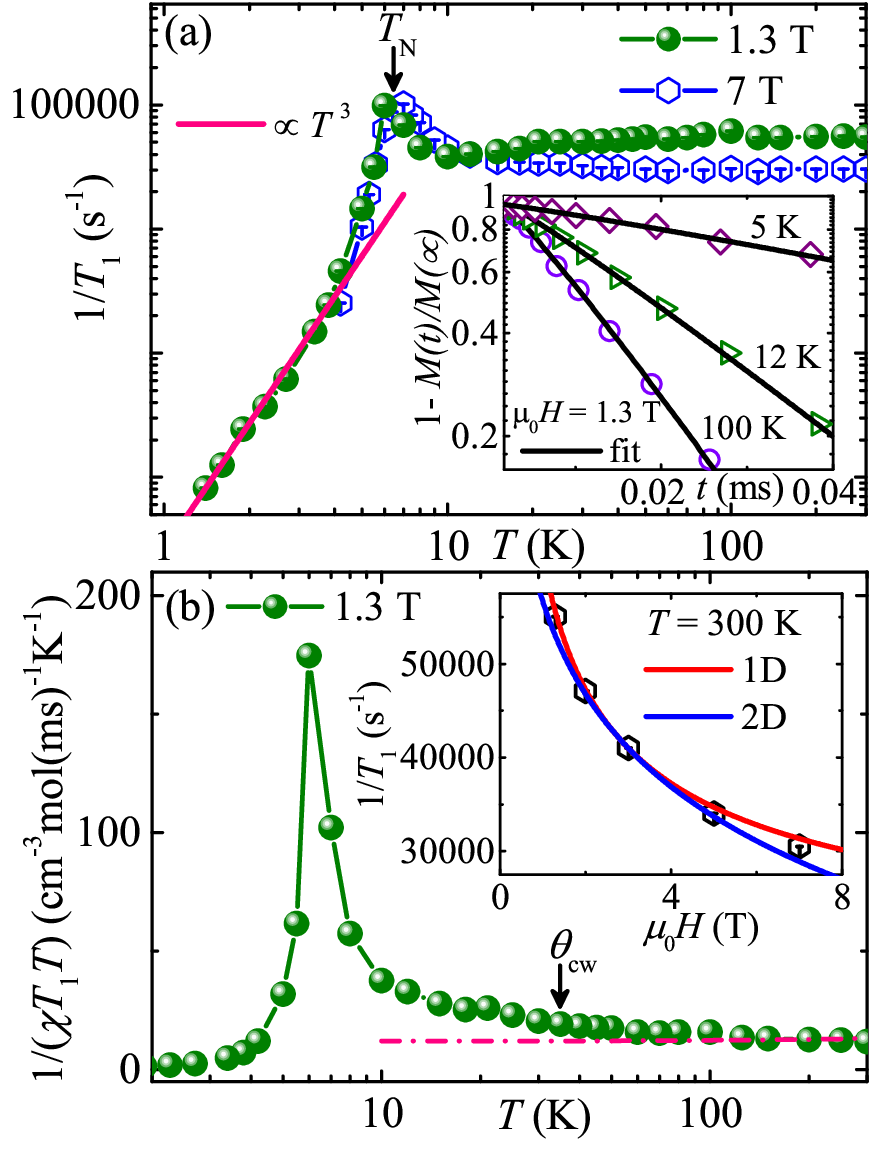}
	\caption{(a) $^{31}$P NMR spin-lattice relaxation rate ($1/T_1$) vs $T$ measured in 1.3~T and 7~T at the P(1) site. The downward arrow points to $T_{\rm N}$. The solid line represents $T^3$ behavior below $T_{\rm N}$. Inset: Longitudinal magnetization recovery curves in 1.3~T at three selective temperatures measured at the P(1) site and the solid lines are fits using Eq.~\eqref{exp}. (b) The temperature dependence of $1/(\chi T_1T)$ measured at 1.3~T. Inset: $1/T_1$ vs $\mu_{\rm 0}H$ measured at 300~K. The solid lines represent the fit using 1D and 2D spin-diffusion models.}
	\label{Fig10}
\end{figure}
To understand the local electron spin-spin correlation, $^{31}$P spin-lattice relaxation rate $1/T_1$ was measured as a function of temperature down to 1.5~K at the central peak position of the P(1) site in two different fields, 1.3~T and 7~T. For an $I = 1/2$ nucleus, the recovery of the longitudinal magnetization is expected to follow a single exponential behavior. Indeed, our recovery curves were fitted well by a stretched exponential function
\begin{equation}
	1-\frac{M(t)}{M(\infty)}= Ae^{-(t/T_{1})^\beta},
	\label{exp}
\end{equation}
where, $M(t)$ is the nuclear magnetization at a time $t$ after the saturation pulse, $M(\infty)$ is the equilibrium nuclear magnetization, and $\beta$ is the stretch exponent. Recovery curves for 1.3~T in three different temperatures along with the fits are shown in the inset of Fig.~\ref{Fig10}(a).

The temperature dependence of $1/T_1$ extracted using the above fitting procedure for the P(1) site are presented in Fig.~\ref{Fig10}(a) for both the fields. Measurements at few selective temperatures at the P(2) site also give same $1/T_1$ as that of the P(1) site as expected, since both the sites have nearly same hyperfine couplings. $1/T_1$ in the high temperature regime ($T>30$~K) is almost temperature independent due to the localized moments in the paramagnetic state. At low-temperatures, $1/T_1$ exhibit a sharp peak at around $T_{\rm N}\simeq 6$ and 7~K for 1.3~T and 7~T, respectively, indicating the slowing down of fluctuating moments as we approach the magnetic ordering. These values of $T_{\rm N}$ corroborate the transition temperatures detected from the $\chi(T)$ and $C_{\rm p}(T)$ measurements. Below $T_{\rm N}$, $1/T_1$ drops swiftly toward zero because of the release of critical fluctuations and the scattering of magnons by the nuclear spins.
For $T > \Delta/k_{\rm B}$, $1/T_1$ should follow either a $T^3$ behavior or a $T^5$ behavior due to a two-magnon Raman process or a three-magnon process, respectively, where $\Delta/k_{\rm B}$ is the energy gap in the spin-wave excitation spectrum. However, at very low temperatures (i.e. $T < \Delta/k_{\rm B}$), it should follow an activated behavior, $1/T_1 \propto T^2e^{-\Delta/k_{\rm B}T}$, due to the opening of a gap in the magnon spectrum. As shown in Fig.~\ref{Fig10}(a), $1/T_1$ below $T_{\rm N}$ follows a nearly $T^3$ behaviour, ascertaining that the relaxation is mainly governed by a two magnon process similar to that reported for other frustrated magnets~\cite{Mohanty104424,Devi012803}.

In Fig.~\ref{Fig10}(b), $1/(\chi T_1T)$ is plotted against temperature. At high temperatures, it is almost temperature independent and then increases slowly below about 40~K. The general expression for $\frac{1}{T_1T}$ in terms of the dynamic susceptibility $\chi_M(\vec{q},\omega_{N})$ can be written as~\cite{Moriya516}
\begin{equation}
\frac{1}{T_{1}T} = \frac{2\gamma_{N}^{2}k_{B}}{N_{\rm A}^{2}}
\sum\limits_{\vec{q}}\mid A(\vec{q})\mid
^{2}\frac{\chi^{''}_{M}(\vec{q},\omega_{N})}{\omega_{N}}.
\label{t1form}
\end{equation}
Here, the sum is over the wave vector $\vec{q}$ within the first Brillouin zone, $A(\vec{q})$ is the form-factor of the hyperfine interaction, and $\chi^{''}_{M}(\vec{q},\omega _{N})$ is the imaginary part of the dynamic susceptibility at the nuclear Larmor frequency $\omega _{N}$. For $q=0$ and $\omega_{N}=0$, the real component of $\chi_{M}(\vec{q},\omega_{N})$ represents the uniform static susceptibility ($\chi$). Thus, the temperature-independent $1/(\chi T_1T)$ in the high temperature region ($T > \theta_{\rm CW}$) indicates the dominant contribution of $\chi$ to $1/T_1T$. The slow increase below $\sim 40$~K reflects the growth of spin fluctuations with $q \neq 0$ or AFM correlations, as expected for a frustrated low-dimensional spin system~\cite{Nath214430}. The steep increase in $1/(\chi T_1T)$ below 10~K is obvious because of the onset of magnetic LRO.

From the constant value of $1/T_1$ at high-temperatures, one can estimate the leading AFM exchange coupling between Fe$^{3+}$ ions. At high temperatures, $1/T_1$ can be expressed as
\begin{equation}
\left(\frac{1}{T_1}\right)_{T\rightarrow\infty} =
\frac{(\gamma_{N} g\mu_{\rm B})^{2}\sqrt{2\pi}z^\prime S(S+1)}{3\,\omega_{ex}}
\left(\frac{A_{z}}{z^\prime}\right)^{2},
\label{t1inf}
\end{equation}
where $\omega_{ex}=\left(|J^{\rm max}|k_{\rm B}/\hbar\right)\sqrt{2zS(S+1)/3}$ is the Heisenberg exchange frequency, $z$ is the number of nearest-neighbor spins of each Fe$^{3+}$ ion, and $z^\prime$ is the number of nearest-neighbor Fe$^{3+}$ spins attached to a given P site. In NH$_{4}$Fe(PO$_{3}$F)$_2$, each Fe$^{3+}$ ion in the triangular plane has six nearest-neighbours. Similarly, each P site is strongly connected to three nearest-neighbour Fe$^{3+}$ spins. Thus, using the parameters $A_{\rm z} \simeq 0.56$~T/$\mu_{\rm B}$, $\gamma_N = 108.303 \times 10^2\,{\rm rad}$~sec$^{-1}$\,Oe$^{-1}$, $z=6$, $z^\prime=3$, $g=2.12$, $S=\frac52$, and the relaxation rate at 125~K $\left(\frac{1}{T_1}\right)_{T\rightarrow\infty}\simeq 54495.91$ sec$^{-1}$, the magnitude of the leading AFM exchange coupling is calculated to be $J^{\rm max}/k_{\rm B}\simeq 1$~K which is the same order of magnitude as $J/k_{\rm B}$ extracted from static $\chi(T)$.

In order to assess the effect of spin diffusion, we plotted $1/T_1$ against $\mu_{\rm 0}H$ measured at $T=300$~K in the inset of Fig.~\ref{Fig10}(b). $1/T_1$ decreases with increase in the magnetic field. It is known that diffusive spin dynamics is observed in low-dimensional Heisenberg magnets for long-wavelength ($q= 0$) spin fluctuations. This is because of the divergence behavior of the spectral density of the spin-spin correlation as $\omega$ tends to zero and this depends on the dimensionality of the spin lattice. Due to spin diffusion, $1/T_1$ is expected to show magnetic field dependence and follows $1/\sqrt{H}$ and $log (1/H$) behaviors for one-dimensional (1D) and 2D systems, respectively~\cite{Takigawa4612,Yogi024413,Ajiro420}. As evident from the inset of Fig.~\ref{Fig10}(b), the experimental data fit well to both the functions reflecting diffusive dynamics. However, it fails to distinguish the nature (1D or 2D) of fluctuations.

\subsubsection{Spin-spin relaxation rate $1/T_2$}
\begin{figure}
	\includegraphics[width=\linewidth]{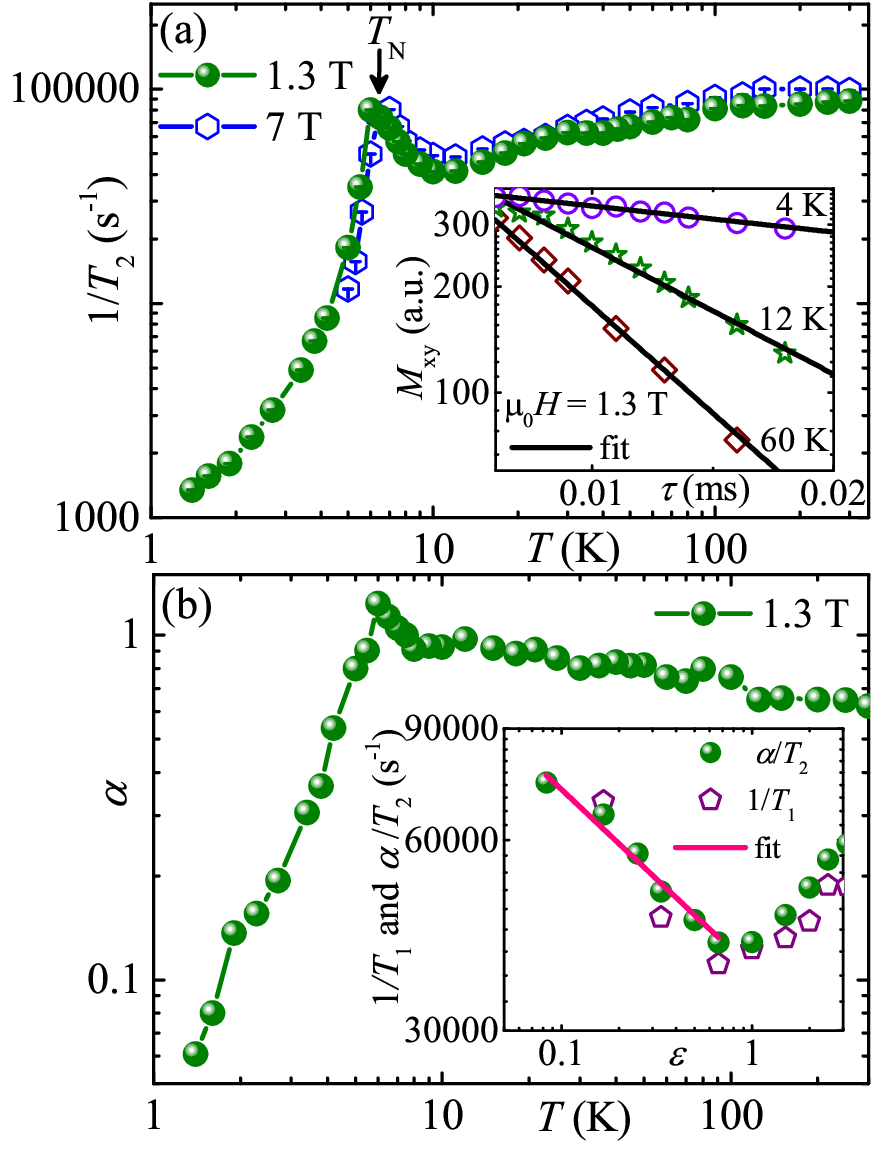}
	\caption{(a) $^{31}$P NMR spin-spin relaxation rate ($1/T_2$) vs $T$ measured in 1.3~T and 7~T at the P(1) site. The downward arrow points to $T_{\rm N}$. Inset: Transverse magnetization recovery curves as a function of $\tau$ at three different temperatures measured at the P(1) site. The solid lines are the fits using Eq.~\eqref{exp2}. (b) Temperature dependence of $\alpha~[=(1/T_1)/(1/T_2)]$ for 1.3~T. Inset: $1/T_1$ and $\alpha/T_2$ vs reduced temperature ($\epsilon$). The solid line is a power law fit.}
	\label{Fig11}
\end{figure}
In order to measure the spin-spin relaxation rate $1/T_2$, the decay of the transverse magnetization
($M_{\rm xy}$) was monitored after a $\pi/2$ - $\tau$ - $\pi$ pulse sequence as a function of the pulse separation time $\tau$. The recovery curves were then fitted by the following equation
\begin{equation}
	M_{\rm xy}= M_0 e^{-(2\tau/T_{2})}.
	\label{exp2}
\end{equation}
Recovery curves at three selected temperatures along with the fits are depicted in the inset of Fig.~\ref{Fig11}(a). The extracted $1/T_2$ in 1.3~T and 7~T are plotted as a function of temperature in Fig.~\ref{Fig11}(a). $1/T_2$ is almost temperature independent at high temperatures, decreases slowly as we lower the temperature, exhibits a sharp anomaly at $T_{\rm N} \simeq 6$~K and 7~K for 1.3 and 7~T, respectively, and then decreases rapidly below $T_{\rm N}$. The overall temperature dependent behaviour of $1/T_2$ is similar to $1/T_1$, though there is a difference in the absolute values. 

For a $I = 1/2$ system, $1/T_2$ is related to $1/T_1$ as
\begin{equation}
	\frac{1}{T_2} = \left(\frac{1}{T_2}\right)^* + \left(\frac{1}{2T_1}\right) + F_z (0),
	\label{exp3}
\end{equation}
where, $1/T_1$ = $F_\perp(\omega_{\rm N}$). $F_z$ and $F_\perp$ are the spectral density of the longitudinal and transverse components of the fluctuating local field, respectively. ($1/T_2$)$^*$ is temperature independent and originates from nuclear dipole-dipole interaction~\cite{Slichter1989,Abragam1961,Devi012803}. For 1.3~T, $1/T_2$ at very low temperatures in the AFM ordered state is nearly temperature independent and can be ascribed to this contribution. Using the lowest value of $1/T_2\simeq1343$~sec$^{-1}$ (at $T = 1.6$~K), the spectral width is calculated to be $\sim 0.77$~Oe, which is of the order of nuclear dipolar field at the P(1) site. Thus, the temperature dependence of $1/T_2$ is originating from $F_z (0)$ and $F_\perp(\omega_{\rm N}$). In order to check this, $\alpha~[=(1/T_1)/(1/T_2)]$ is plotted as a function of temperature in Fig.~\ref{Fig11}(b). $\alpha$ is almost constant above $\sim 40$~K and increases with decreasing temperature and becomes $\alpha \simeq 1.1$ at $T_{\rm N}\simeq 6$~K. Below $T_{\rm N}$, it decreases rapidly due to the AFM ordering. For $\alpha\simeq0.7$ above 40~K (in the paramagnetic state), we get $F_z (0)\simeq 0.92~F_\perp(\omega_{\rm N}$), which suggests that $F_z (0)$ and $F_\perp(\omega_{\rm N}$) contribute equally to $1/T_2$.

Furthermore, at the ordering temperature, the correlation length is expected to diverge and $1/T_1$ in a narrow temperature range just above $T_{\rm N}$ (i.e., in the critical regime) should be described by a power law, $1/T_1 \propto \epsilon^{-\gamma}$, where $\gamma$ is the critical exponent and $\epsilon = (T-T_{\rm N})/T_{\rm N}$ is the reduced temperature. The value of $\gamma$ characterizes the universality class of the spin system depending upon its dimensionality, symmetry of the spin lattice, and the type of interactions. To analyze the critical behavior, both $1/T_1$ and $\alpha/T_2$ (taking $\alpha\simeq1.1$ near $T_{\rm N}$) are plotted against $\epsilon$ in the inset of Fig.~\ref{Fig11}(b). The data just above $T_{\rm N}$ ($\epsilon \leq 0.7$) were fitted by the power law with a fixed $T_{\rm N}\simeq 6$~K that yields $\gamma \simeq 0.28$. For a 3D Heisenberg antiferromagnet, the mean-field theory predicts $\gamma = 1/2$ while the dynamic scaling theory gives $\gamma = 1/3$~\cite{Lee214416}. Our experimental value $\gamma \simeq 0.28$ is close to 0.3 expected for a 3D Heisenberg spin system but far below 0.8 for the 2D Heisenberg spin system, indicating that the AFM ordering is driven by 3D correlations~\cite{Benner1990,Ranjith024422}.


\subsection{Phase Diagram}
\begin{figure}
\includegraphics[width=\columnwidth]{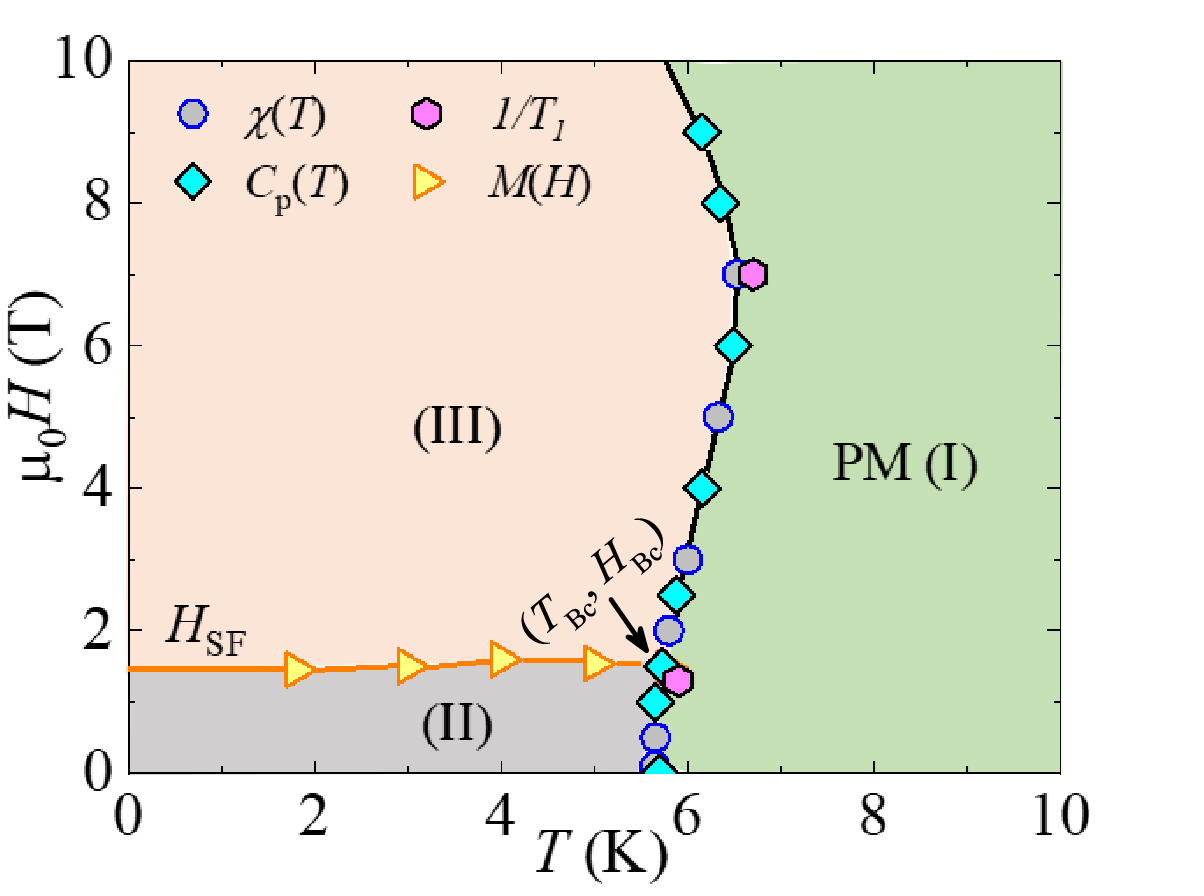}	
\caption{$H-T$ phase diagram obtained from the magnetic isotherm, susceptibility, heat capacity, and NMR spin-lattice relaxation data. The shaded regions represent three distinct phases (I, II, and III). The point ($T_{\rm Bc}, H_{\rm Bc})$ is the bicritical point, where the SF transition ends.}
\label{Fig12}
\end{figure}
The $H - T$ phase diagram constructed using $T_{\rm N}$ values obtained from $\chi(T)$, $C_{\rm p}(T)$, and $^{31}1/T_1$ along with $H_{\rm SF}$ corresponding to the metastable transition observed from the magnetic isotherms is presented in Fig.~\ref{Fig12}. It features three distinct regions: paramagnetic (I), antiferromagnetic (II), and spin-flop (III) phases~\cite{Sebastian104425,Kikuchi012117}. In case of a conventional antiferromagnet, $T_{\rm N}$ should move progressively towards lower temperatures with increasing field. However, in our case, the phase boundary moves towards higher temperatures with field upto $\sim 6$~T and then towards low temperatures in higher fields. This shape of the phase boundary doesn't reflect a simple antiferromagnet.
This type of phase boundary is often observed in canted antiferromagnets where magnetic anisotropy induces spin canting~\cite{Garlea011038,Lee224402,Povarov214402}. Secondly, this shape can also be explained in terms of competing AFM LRO and quantum fluctuations. In low-dimensional and frustrated magnets, quantum fluctuations suppress the magnetic LRO. External magnetic field weakens these fluctuations, therefore $T_{\rm N}$ is initially enhanced in low fields. Upon further increase in field, the tendency towards the fully polarized state conquers the AFM state, hence, $T_{\rm N}$ moves towards low temperatures~\cite{Nath064422,Tsirlin014429,Kohama184402}.

The true ground state of a Heisenberg TLAF is a non-collinear $120^{\degree}$ order~\cite{Capriotti3899}. When the magnetic field is applied, the system evolves from a $120^{\degree}$ Néel order to a $uud$ state and manifests as a magnetization plateau at $1/3$ of the saturation magnetization in the magnetic isotherms~\cite{Chubukov69,Ono104431,Susuki267201,Smirnov134412}.
This $uud$ phase is stabilized by the axial (Ising) anisotropy induced by the magnetic field.
Though the magnetic isotherms of NH$_4$Fe(PO$_3$F)$_2$ feature a plateau in intermediate fields but it is well below $1/3$ of the saturation magnetization, ruling out the $1/3$ plateau~\cite{Seabra214418}. 
Further, the low temperature NMR spectra divulges a commensurate/collinear AFM order below $T_{\rm N}$ in zero-field, in contrast to an expected non-collinear $120^{\degree}$ order~\cite{Sebastian104425}. This can be attributed to the distortion in FeO$_6$ octahedra and inter-layer coupling along the stacking $c$-direction~\cite{Tapp064404,Sebastian104425}. Moreover, it is reported that the $uud$ phase is absent in isotropic or easy-plane (XY-type) anisotropy systems and only single transition survives in zero-field. Such systems often show the meta-magnetic transitions like SF transition above a critical field $H_{\rm SF}$, where the moments flip from parallel to perpendicular direction with respect to the applied field~\cite{Lee224402,Smirnov134412}. Thus, the observed metastable SF transition can be ascribed to the easy-plane anisotropy in NH$_4$Fe(PO$_3$F)$_2$. Indeed, other TLAFs like Na$_3$Fe(PO$_4$)$_2$, Cu$_2$(NO$_3$)(OH)$_3$ and Na$_2$BaMnV$_2$O$_8$ also show a SF transition at low temperatures~\cite{Nakayama116003,Sebastian104425,Kikuchi012117}. Similar phase diagram is also reported previously in the frustrated and anisotropic spin chain compounds SrCuTe$_2$O$_6$ and $\alpha$-Cu$_2$As$_2$O$_7$~\cite{Ahmed214413,*Arango134430}.


\section{SUMMARY}
In summary, we have performed a detail study and examined the magnetic ground state of a distorted TLAF NH$_{4}$Fe(PO$_{3}$F)$_2$. In the crystal structure, the 2D triangular layers are slightly buckled yielding a small anisotropy in the triangular units. Despite two inequivalent P-sites in the crystal structure, the $^{31}$P NMR reveals nearly identical hyperfine couplings ($A_{\rm hf}^{\rm P(1)} \simeq 0.56$~T/$\mu_{\rm B}$ and $A_{\rm hf}^{\rm P(2)} \simeq 0.62$~T/$\mu_{\rm B}$) for both the P-sites with the Fe$^{3+}$ spins. The analysis of $\chi(T)$ and $K(T)$ establishes that the system behaves like a spin-$5/2$ isotropic triangular lattice with an average NN exchange coupling $J/k_{\rm B} \simeq 1.7$~K. The observed anomaly in $\chi(T)$, $C_{\rm p}(T)$, $1/T_1$, and $1/T_2$ and the instant NMR line broadening pin point the transition to a LRO state at $T_{\rm N} \simeq 5.7$~K. Further, the critical analysis of relaxation rates demonstrates that the AFM LRO is driven by 3D correlations. The field induced SF transition in low temperatures implies the presence of XY-type anisotropy. $^{31}$P NMR spectra below $T_{\rm N}$ unfolds the nature of the ordering to be commensurate AFM type which is stabilized possibly due to either distortion in the FeO$_6$ octahedra or inter-layer coupling. At the end, the $H-T$ phase diagram is discussed.

\acknowledgments
We would like to acknowledge SERB, India for financial support bearing sanction Grant No.~CRG/2022/000997. We also acknowledge the support of HLD-HZDR, member of the European Magnetic Field Laboratory (EMFL). KMR acknowledges the financial support by the Deutsche Forschungsgemeinschaft (DFG, German Research Foundation) under Germany’s Excellence Strategy through the Würzburg-Dresden Cluster of Excellence on Complexity and Topology in Quantum Matter-ct.qmat (EXC 2147, project-id 390858490).

%

\end{document}